\documentclass{sig-alternate}

\usepackage{url}

\newtheorem{theorem}{Theorem}
\newtheorem{defn}{Definition}

\newcommand{\onperiod}{{\em on\/} period\,}
\newcommand{\onperiods}{{\em on\/} periods\,}
\newcommand{\offperiod}{{\em off\/} period\,}
\newcommand{\offperiods}{{\em off\/} periods\,}
\newcommand{\mr}{\mathbb{R}} 
\newcommand{\mn}{\mathbb{N}} 

\newcommand{\Esym}{\mathrm{E}}
\newcommand{\E}[1]{\Esym\left[#1\right]}
\newcommand{\Probsym}{\mathbb{P}}
\newcommand{\Prob}[1]{\Probsym\left[#1\right]}

\begin{document}

%

\title{Criticisms of modelling packet traffic using long-range dependence (Extended Version)}

\numberofauthors{1} %
\author{
\alignauthor Richard G. Clegg, Raul Landa, Miguel Rio \\
       \affaddr{Dept of Electrical \& Electronic Engineering}\\
       \affaddr{University College London, UK}\\
       \email{ \{rclegg,r.landa,m.rio\}@ee.ucl.ac.uk}
}

\date{20 November 2009}

\additionalauthors{}

\maketitle
\begin{abstract}
This paper criticises the notion that long-range dependence is an important
contributor to the queuing behaviour of real Internet traffic.  The idea
is questioned in two different ways.  Firstly, a class
of models used to simulate Internet traffic is shown to have important
theoretical flaws.  It is shown that this behaviour is inconsistent with
the behaviour of real traffic traces.  Secondly, the notion that long-range
correlations significantly affects the queuing performance of traffic
is investigated by destroying those correlations in real traffic traces 
(by reordering).  It is shown that the longer ranges of correlations are
not important except in one case with an extremely high load.
\end{abstract}

\section{Introduction}
\label{sec:intro}

Since the seminal paper of  Leland et al 
\cite{leland1993} it has been
considered important that a statistical model of Internet traffic
captures the phenomenon of Long-Range Dependence (LRD).  In particular
it has often been suggested that a model of Internet traffic must capture
the Hurst parameter $H \in (1/2,1)$ of real traffic.
LRD is characterised by the unsummability of the
autocorrelation function (ACF).  It is often stated that this is an 
important characteristic for the queuing performance of the traffic.

A related topic is that of heavy-tailed distributions.  A commonly suggested origin
for the LRD in Internet traffic is the heavy-tailed distribution of traffic
\onperiods.
\begin{defn}
A random variable $X$ is heavy-tailed if, for all $\varepsilon > 0$
it satisfies
\begin{equation}
\Prob{X > x} e^{\varepsilon x} \rightarrow \infty \quad \mathrm{ as } \quad 
x \rightarrow \infty.
\label{eqn:exptails}
\end{equation}
\end{defn}
A specific functional form is usually assumed (and will be throughout this paper)
\begin{equation}
\Prob{X > x} \sim C x^{-\beta},
\label{eqn:tails}
\end{equation}
where $C > 0$ is a constant and $2 > \beta > 0$.  The symbol $\sim$ means
asymptotically equal to.  If $\beta < 1$ then
$\E{X}$ is infinite and therefore most models use $\beta \in (1,2)$.

Suggested models for Internet traffic which generate LRD include
fractional Gaussian noise and the related
fractional Brownian motion (fGn, fBm) 
\cite{paxson1997}, chaotic maps
\cite{erramilli1994}, wavelets \cite{riedi1999,riedi2003}
and Markov modulated processes \cite{barenco2004,clegg2005}.  Some
of these models output a ``traffic level" which represents the mean
arrival rate in some notional time period but others are packet
based models, that is they produce a model of packets and inter-arrival
times.  It is the latter class of models (including 
\cite{erramilli1994,barenco2004,clegg2005} which are covered by theorem \ref{thm:1}
in this paper.

This paper criticises the notion that the long-range correlations in traffic
are important to queuing in two ways.  In section \ref{sec:theory} it is
shown that a class of models used to simulate traffic with LRD arising
from heavy tails gives an infinite result when queued in infinite buffers.
Section \ref{sec:framework} describes the simulation framework
used for the rest of the paper.
It is demonstrated in section \ref{sec:traffheavy} that this is at odds
with the behaviour of real traffic.  In section \ref{sec:reorder} real
traffic traces are analysed again and reordered to break up correlations
beyond a certain level.  It is shown that this reordering does not affect
the queueing behaviour of the traffic beyond a certain time-scale except
when unrealistically high loads are used.  The behaviour of the long-range
dependent models (and in particular a certain class based on heavy-tails)
is theoretically undesirable and fundamentally different to that of real
traffic.

\section{Theoretical results}
\label{sec:theory}

Let $\{A_t: t \geq 0 \}$ be an arrival process to a queue drained
by a deterministic server which serves at a constant rate assumed
without loss of generality to be one.  The mean arrival rate $\lambda$ is 
given by $\lambda= \lim_{T \rightarrow \infty} \int_{0}^{T} A(t) dt /T$
and it is assumed throughout that $A_t$ is such that this limit exists
and $\lambda \in (0,1)$.  Since the server rate is one then $\lambda$
is equal to the utilisation $\rho$ (the ratio of the
rate at which work enters to the maximum rate at which it can be served).
Let $\{Q_t: t \geq 0 \}$ be the queue process where it is assumed that
$Q_0 = 0$.  Assume that the queue evolves according to 
$$\frac{dQ_t}{dt} =
\begin{cases}
A_t - 1 & Q_t > 0 \\
\max(0, A_t -1) & Q_t = 0.
\end{cases}
$$
Let $\E{Q(s,t)} = \int_s^t \E{Q_u} du/(t-s)$
where $t > s$ and $\E{\cdot}$ denotes expectation.  Let
$\E{Q} = \lim_{t \rightarrow \infty} \E{Q(0,t)}$ and
note that this  limit is not guaranteed to exist (and may tend to
infinity).
The mean arrival rate at time $t$ is 
$\lambda_t = \E{A_t}$ and the overall
mean arrival rate $\lambda= \lim_{T \rightarrow \infty}
\int_0^T \lambda_t/T dt$. 
If $\lambda > 1$ then $\rho > 1$ and
the queue must eventually grow to infinity regardless of the details of
the arrival
process.

\begin{theorem} \label{thm:1}
Let $\{A_t: t \in \mr_+\}$ be an ergodic, weakly stationary 
arrival process.  
This is drained by a queue which drains at a fixed rate $r$.  
Let $A_t$ be such that
the utilisation $\rho$ is in $(0,1)$.  Call $A_t$ on
when $A_t \geq m$ (where $m$ is any constant such that
$m > r$) and off otherwise.
Let $\{X_n: n \in \mn\}$ be the length of 
the $n$th \onperiod.  If the $X_n$
are i.i.d. with a heavy-tailed
distribution $\Prob{X_n > x} \sim x^{-\alpha}$ for 
$\alpha \in (1,2)$ then as $t \rightarrow
\infty$ the expected
queue length (and hence the expected waiting time) is infinite.
\end{theorem}

This theorem can be restated as: if an infinite buffer queuing model is driven 
by a single on/off source with the utilisation in
$0,1$, i.i.d. heavy-tailed \onperiods
then the expected queue length is either zero (if $m \leq r$) or
infinite.  This remains true even if the utilisation is arbitrarily close
to zero and the amount by which the demand exceeds the queue drain rate
$m - r$ is arbitrarily small.
It may seem paradoxical that a queue which is empty arbitrarily often
can have an infinite expected length.  However, this has parallels with
the classical Pollaczek--Khinchine formula for an M/G/1 queue
\cite{khinchin1932} where
a server with an infinite variance in the service time has an infinite
expected queue length even if the mean service time is arbitrarily small
and the queue empty an arbitrarily large proportion of the time.

Note that for $\alpha < 1$ the mean length of an \onperiod will not converge,
the utilisation will not be in the range $(0,1)$ and
such processes will not, in general, be useful for a queuing system.  However,
for $1 < \alpha < 2$ the mean length of an \onperiod will be finite and such a
process could be used to produce a time series with a known Hurst parameter.

The strategy of the proof below is to find a simple process $A'_t$
which gives a lower bound on the queuing delay for any $A_t$ meeting
the conditions of the theorem.  It is then shown that the mean
queue length and hence mean delay for $A'_t$ is infinite.

\begin{proof}
Without loss of generality, let $r=1$.  In this case the arrival rate 
$\lambda$ is equal to the utilisation $\rho$.  Assume 
$A_t$ is such that $A_t = m$ during an
\onperiod and $A_t = 0$ during an \offperiod (such a process will
always have a smaller queue than the specified process where $A_t > m$ in an
\onperiod and $A_t \in [0,a)$ and thus can be considered a lower bound).

First consider a single \onperiod followed by a single \offperiod of such
length that the entire queue has drained by the end of the \offperiod.
Consider the time period $(t_1,t_2)$ where $Q_{t_1} = Q_{t_2} = 0$,
consisting of an
on period $(t_1, t_1+X)$ (where $mX < (t_2 - t_1)$)
and an \offperiod $(t_1+X,t_2)$.  Within the period $(t_1,t_2)$
the queue peaks at time
$t_1 + X$ when $Q_{t_1+X} = (m-1)X$ and drains
completely by time $t_1+mX$ after which the queue is zero until $t_2$.
It can be readily seen that, since the queuing process is triangular in
shape (rising at rate $m-1$ during the \onperiod and falling at rate 1
during the \offperiod), then
$\int_{t_1}^{t_2} \E{Q_u} du = (m-1)mX^2/2$ and 
$\E{Q(t_1,t_2)} = (m-1)mX^2/2(t_2-t_1)$.

Now consider some time period $(t_1, t_2)$ again where $Q_{t_1} = 
Q_{t_2} = 0$.  Let this period contain exactly two
\onperiods of lengths $X_1$ and $X_2$ where $m(X_1 + X_2) < (t_2 - t_1)$. 
It is clear that 
$\int_{t_1}^{t_2} \E{Q_u} du\geq (m-1)m(X_1+X_2)/2$  with equality occurring
only when the queue empties completely between the two \onperiods.
This argument can be trivially extended  to
$n$ \onperiods
of lengths $X_1, X_2, \ldots, X_n$ 
all occurring within $(t_1,t_2)$ with $Q_{t_1} = Q_{t_2} = 0$.  The mean queue
size is minimised if the \onperiods are such that the generated queues do
not overlap.

Consider the process $A'_t$ 
which has the same mean arrival rate and
is the process $A_t$ reordered in time according to the following rules:
\begin{itemize}
\item \onperiods occur in the same order and have the same length
as $A_t$ with the first \onperiod starting
at $t=0$,
\item an \onperiod of length $X_i$ is followed by an \offperiod of length 
exactly $X_i(m/\lambda - 1)$.
\end{itemize} 
This \offperiod is long enough that the queue has always completely drained
before the end of the \offperiod (since $\lambda < 1$).  It can easily be shown 
that such a reordering is possible since the \onperiods are of exactly the same
length in the same order and the \offperiods have the same mean length.

Let $Q'_t$ be
the queue process for $A'_t$ (assuming the same server process).  Clearly
$\E{Q'} \leq \E{Q}$ since the queues due to $A'_t$ never overlap (with
equality occurring only when the queues never overlap in $A_t$ either).

It can be shown that 
\begin{align*}
\E{Q'} & = \lim_{N \rightarrow \infty}
\frac{\sum_{i=1}^N \int_{t_i}^{t_{i+1}}  Q'_t dt} {t_{N+1}} \\
&=
\lim_{N \rightarrow \infty}
\frac{a(a-1) \sum_{i=1}^N X_i^2 } {2\sum_{j=1}^N X_j}.
\end{align*}
Taking expectations a second time gives
\begin{align*}
\E{\E{Q'}} & = \E{Q'} =  \lim_{N \rightarrow \infty}
\frac{m(m-1) \sum_{i=1}^N \E{X_i^2} } {2\sum_{j=1}^N \E{X_j}}
\\ & = 
\frac{m(m-1) \E{X^2}}{2\E{X}},
\end{align*}
where the last equality follows since the $X_i$ are i.i.d.

$\E{Q'}$ is a lower bound for $\E{Q}$.   $\E{Q}$ does not
converge if $\E{X^2}$ does not converge.  
If $1 < \alpha < 2$ then $\E{X}$ is
finite but $\E{X^2}$ is not and the expected queue is infinite.
The result follows.
\end{proof}

A similar result holds for discrete time on-off arrival processes
$\{A_n: n \in \mn\}$ with heavy-tailed \onperiods.  However, 
the result is not
true, for example, if the queue is driven by two or more heavy-tailed sources
each of which has an arrival rate less than one but together having
an arrival rate over one.  Processes such as fractional Gaussian noise exhibit 
LRD but have a finite expected queue length in an infinite buffer.  In fact
the theorem has an obvious corollary.

It should be noted that heavy-tailed arrival processes of the
type from Theorem \ref{thm:1} which give rise to
a finite value of $\E{Q}$ are possible but paradoxically these have an arrival
rate $\lambda=0$.  For example, assume the queue drains at rate $r$,
let $q > 0$ and let every \onperiod of 
length $X_i$ (with arrivals at exactly rate $m$) be
followed by an \offperiod of length 
$\max((m-1)X_i,(m-1)mX_i^2/2q -  X_i)$.  The queue drains completely in every
\offperiod and 
the mean queue size for that \onperiod and \offperiod is at most $q$
(with equality attained when $((m-1)mX_i^2/2q -  X_i) \geq (m-1)X_i$) and
therefore $\E{Q} \leq q$.  However, in the heavy-tailed case, $\E{X_i^2}$ does
not converge and hence neither does the expected length of an \offperiod.  This
implies a proportion of time in the \offperiod tending to one and a mean
arrival rate $\lambda$ of zero.

It might still be argued that real traffic traces have this property
but the outcome of infinite queue size is not seen in real life because
they are fed to a finite sized array.  This possibility will be
investigated in section \ref{sec:traffheavy}.

\section{Simulation framework for this paper}
\label{sec:framework}

CAIDA data: This data set is taken from a trace approximately an
hour long.  It is referred to on the CAIDA website
\footnote{See \url{http://www.caida.org/data/passive/} for more information
about this trace.}
as \\ {\tt 20030424-000000-0-anon.pcap.gz}
and was captured on
the 24th April 2003.  It
was captured on an OC48 link with a rate of 2.45 Gb/s.  
The first 2,320,137 packets (five minutes) are
used in the analysis here.  This trace has a relatively low 
Hurst parameter $H= 0.6$ (see \cite{clegg2007}).

Bellcore data: This is a much studied data set and, while certainly not
representative of modern traffic, it is included as one of the 
original traces from \cite{leland1993}. 
The data here is taken from an August 1989 measurement referred to as
{\tt BC-pAug89.TL}.
The data was collected
on an Ethernet link\footnote{See
\url{http://ita.ee.lbl.gov/html/contrib/BC.html}
for more information about this traffic}.  The first 1,000,000 packets are
studied here.  This trace has a relatively high Hurst parameter $H=0.8$
(see \cite{clegg2007}).

The simulations used in this paper are all based upon an extremely
simple queuing model.  Packets arrive in a FIFO
buffer which never drops
packets.  
The buffer has a given bandwidth $b$ (in bytes/second) -- which can be
adjusted 
to give a specific level of utilisation.  While the absolute level
of the queue changes, the results presented here are not very sensitive to
this parameter.  A
packet of length $l$ bytes takes $l/b$ seconds to depart the queue.
If $Q_t$ is the queue length in packets at time $t$ and the simulation runs
until time $T$
then the mean queue $\overline{Q}$ 
is evaluated
as $\overline{Q} = \int_0^T Q_t/T dt$ (this integral can be evaluated exactly
since  $Q_t$ is a constant between arrivals and departures from the queue).

This simple model has been used by the first author 
to assess several the queuing 
performance of several models which attempt to replicate the statistical
nature of Internet traffic.  This work is reported in \cite{clegg2007}.  
None of the long-range dependence based models replicated the queuing 
performance of the real traffic traces they were tuned to match.   
Such a simple model is, of course, open to much criticism.  It 
does not account for TCP feedback mechanisms.  However, if the 
question being addressed is about an ``open loop" model of traffic
and whether it replicates the characteristics of real traffic then 
we should expect both to have the same behaviour at a queue.

In all the experiments here each data point is constructed by performing
ten repetitions of the experiment.  The error bars represent one standard
deviation each side of the mean.  Because of the nature of long-range
dependent statistics, in simulation experiments (where the data is
truly long-range dependent) the error bars are often 
extremely large.

\section{Simulation results on real versus heavy-tailed traffic}
\label{sec:traffheavy}

The first simulations here consider the theoretical results presents
in section \ref{sec:theory}.  The first results show how the infinite
expected queue size in the model reveals itself in simulation results.
Obviously any experiment with a model of the form in Theorem \ref{thm:1}
will produce a finite value for $\overline{Q}$ the mean queue size.
However, this finite value will increase as the model is run for longer and longer
(up to the numerical accuracy of the model).  The value of $\overline{Q}$ generated
will (in theory at least -- in practice the finite accuracy of computers limits this)
increase as the runtime increases.  The question may be asked if this is true of real
data.

The experiment performed here is to take different sized samples of the real data
and to queue those samples with the model from the previous section.  The behaviour
of the mean queue length versus sample size is investigated.  Experiments on LRD are
notorious for their high (often theoretically infinite) variability.  Here 
ten replications are performed for each size are used and the mean 
plotted.  In addition error bars of the size of the standard deviation (one
standard deviation above and below the mean) are 
added (standard confidence interval techniques are not applicable in the case of
LRD data).

\begin{figure}[ht!]
\begin{center}
\includegraphics[width=7.5cm]{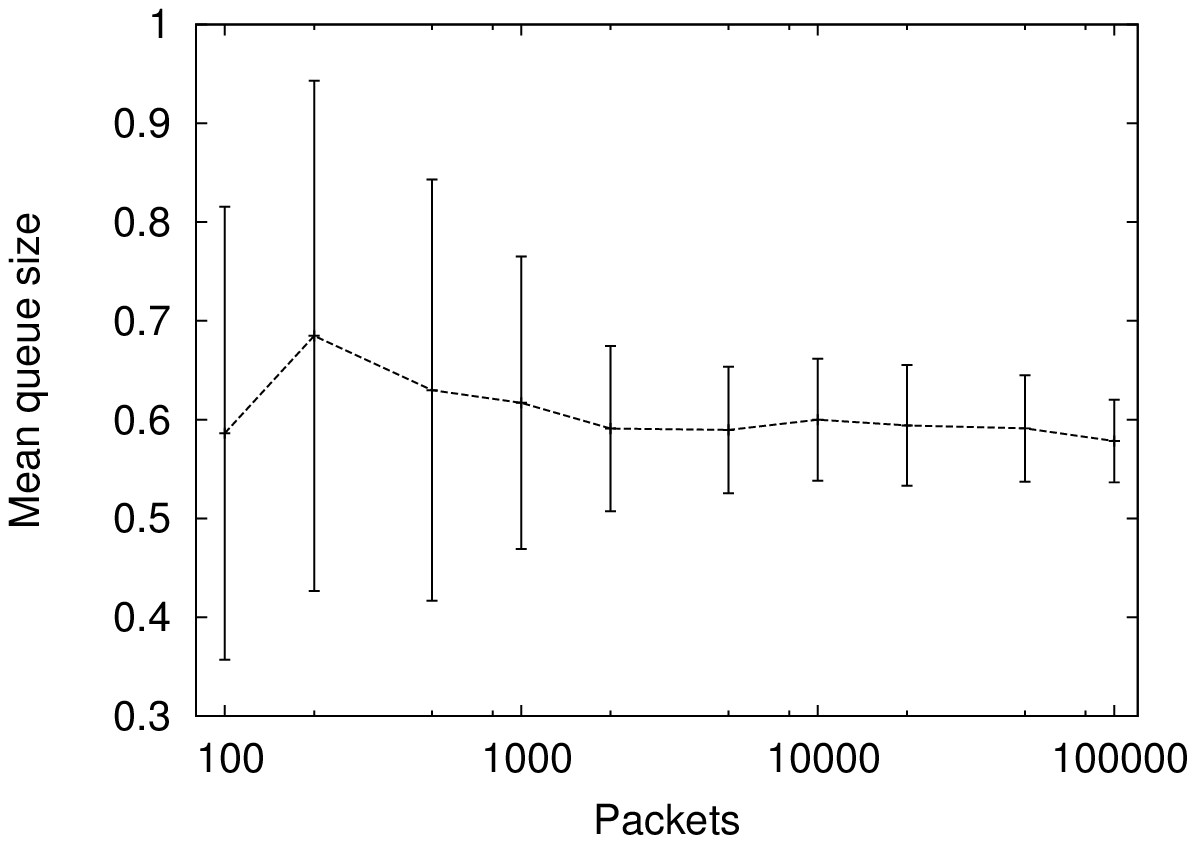} \\
\includegraphics[width=7.5cm]{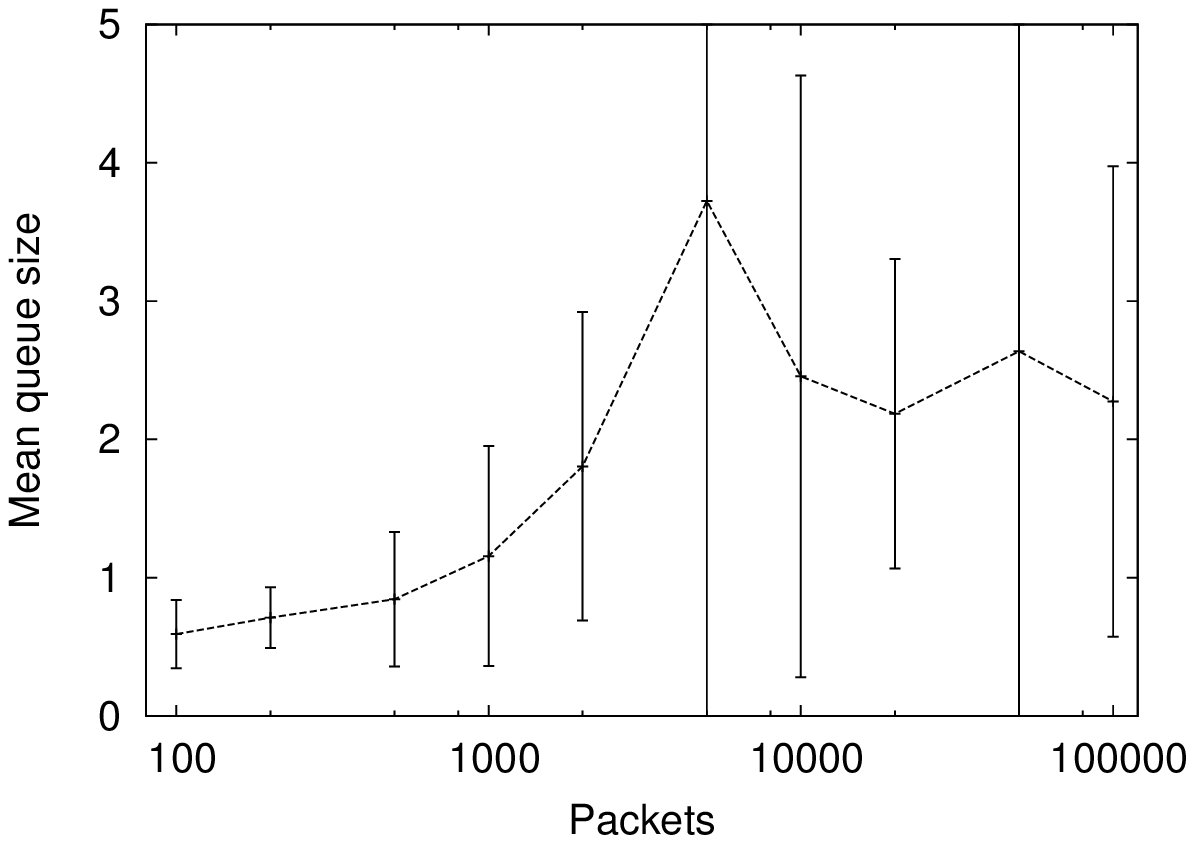}
\caption{Mean queue size versus number of packets for the CAIDA data set --
real data (top) and simulated (bottom).}
\label{fig:oc48plot}
\end{center}
\end{figure}

Figure \ref{fig:oc48plot} (top)
shows the results for the CAIDA data.  The x axis
shows the number of packets in the sample and the y axis the mean queue size
(or rather the mean of the ten means for the ten experiments with that sample
size).
Since the
queue is assumed to start empty then very small sample sizes would naturally
have a smaller expected queue.  However, beyond this, figure \ref{fig:oc48plot}(top)
shows no clear influence of the sample size on the expected queue.  As the samples
get larger the standard deviation bounds from the ten experiments gets smaller. 

Figure \ref{fig:oc48plot} (bottom)
shows the same experiment
but performed on a simulated data set with the same mean arrival rate and the same
Hurst parameter using the techniques described in
\cite{clegg2007} (the Wang model from that paper).  The simulation does not
well reflect the queuing performance of the trace and this is because of
theorem \ref{thm:1} which applies here.  In contrast with figure \ref{fig:oc48plot}
the mean queue size increases with the number of packets in the sample.  In addition
the error bars (representing one standard deviation either side of the mean) stay
as large or become larger.  The increase predicted from theory is somewhat
disguised by the extremely large error bars particularly in the middle part of the
diagram.

\begin{figure}[ht!]
\begin{center}
\includegraphics[width=7.5cm]{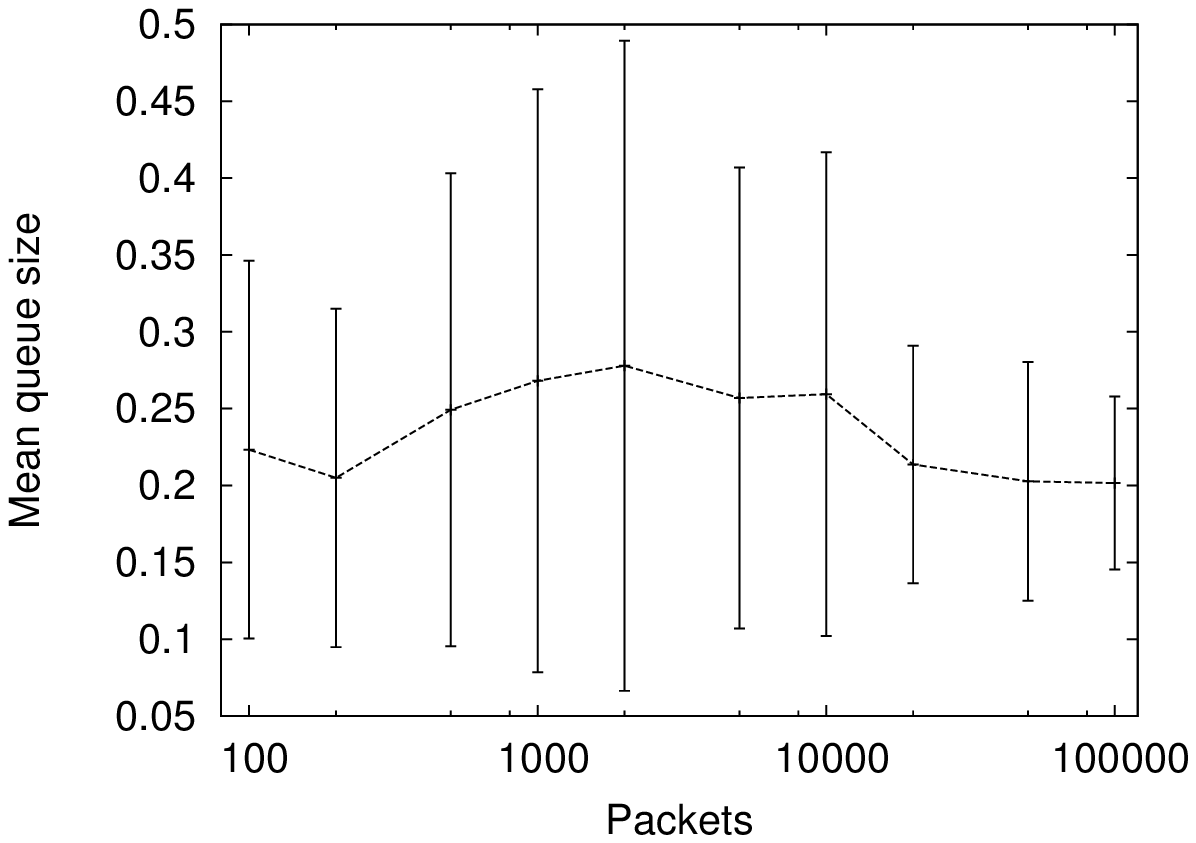} \\
\includegraphics[width=7.5cm]{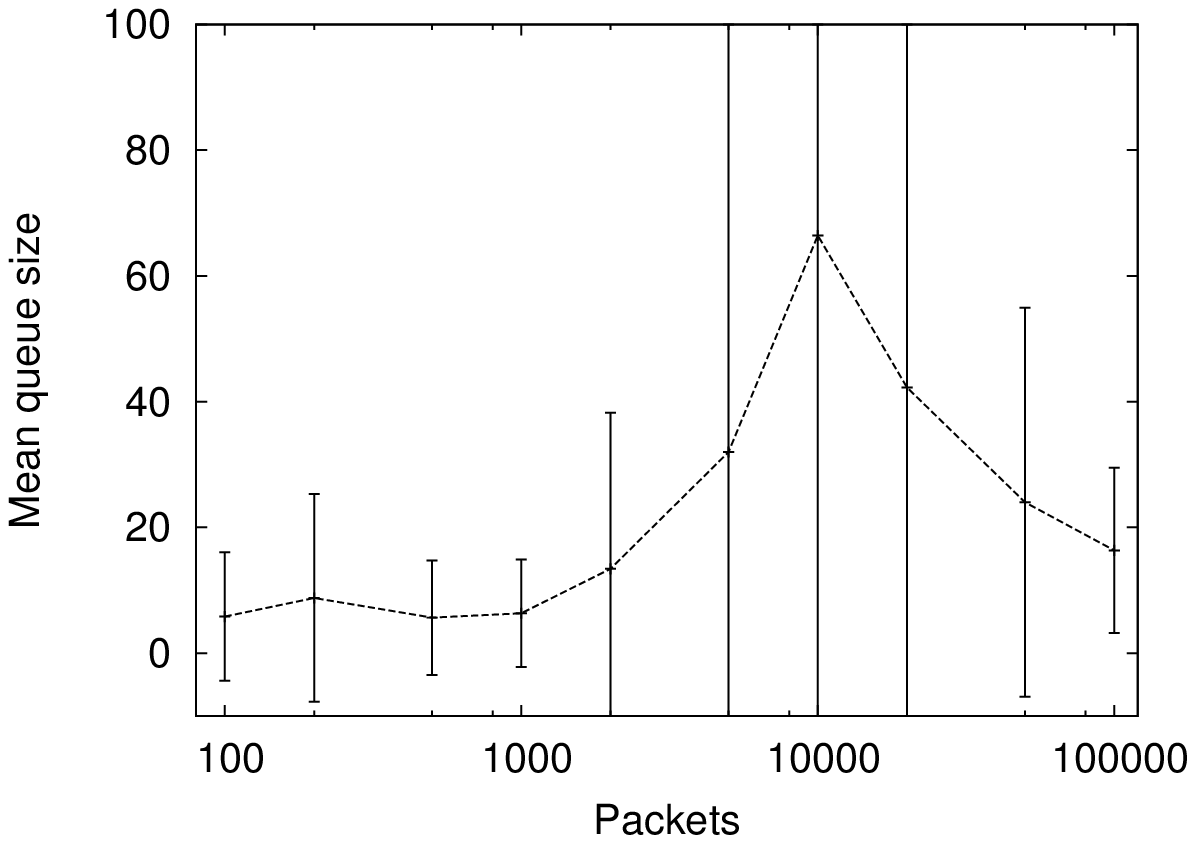}
\caption{Mean queue size versus number of packets for the Bellcore data --
real data (top) and simulated (bottom).}
\label{fig:bc_queue}
\end{center}
\end{figure}

Figure \ref{fig:bc_queue} (top) shows the same experiment for the Bellcore data. 
This data has a higher Hurst parameter.  Again
this figure is not consistent with the idea that the mean queue rises as the
length of the sample rises apart from in the early part of the plot.  (The rise
in the mean and the larger error bars in the center of the plot coincides with
a single very large burst of a particular duration).
Figure \ref{fig:bc_queue} (bottom) shows simulated data with the same Hurst parameter
and same mean arrival rate as
the Bellcore data.  As in figure \ref{fig:oc48plot} (bottom) 
and in accordance with theorem
\ref{thm:1} the mean queue size rises with the number of packets simulated.
The correlation here is again slightly ambiguous because of the large standard deviation
on medium sized samples.  As can be seen, these simulations
can be problematic to work with and a researcher looking only at the early part of
the graphs could be convinced that they had used sufficiently many packets for
the simulation to converge to a good estimation of the mean queue length.

It may be thought that the problem may be connected with the fact that the LRD based
methods dramatically overestimated levels of queuing.  However, repeating the 
experiment with lower bandwidth on the real data for both Bellcore and CAIDA traces
does not dramatically alter the shape of the graph although obviously the mean 
queue level increases.  

To recap then, the results in this section show that for the simulated
data the results are inline with Theorem \ref{thm:1} from which it is
expected that the actual mean queue length is infinite and the finite
queue length is merely a result of the finite experiment length.
This is indicated by the fact that longer
samples have a longer queue length in the data here.  
The results on real data are not
consistent with this behaviour and longer sample sizes do not always
produce a longer queue length.

\section{Simulation results on reordering of real traffic traces}
\label{sec:reorder}

This section takes a different approach by deliberately destroying correlations in
the data to see which scales of correlation are important to the queuing properties.
It is often stated that LRD is an extremely important property for queuing in real data.
If this is the case, then deliberately truncating the correlation beyond a certain
scale should have important effect on the queuing.   

The experiment performed in this section is to take a certain blocksize $B$ and to
split the data into blocks each containing $B$ packets (and associated delays). 
The order of these blocks is then randomised so no correlation can persist beyond 
$B$ packets.  The entire trace is then queued and the mean queue length recorded.
Again ten replications are performed to assess the repeatability.
The important point here is that after reordering, only correlations of less
than the blocksize can possibly remain.  If the correlations in the data are
having an important effect on queuing then breaking those correlations should
have a significant effect on queue size.  

\begin{figure}[ht!]
\begin{center}
\includegraphics[width=7.5cm]{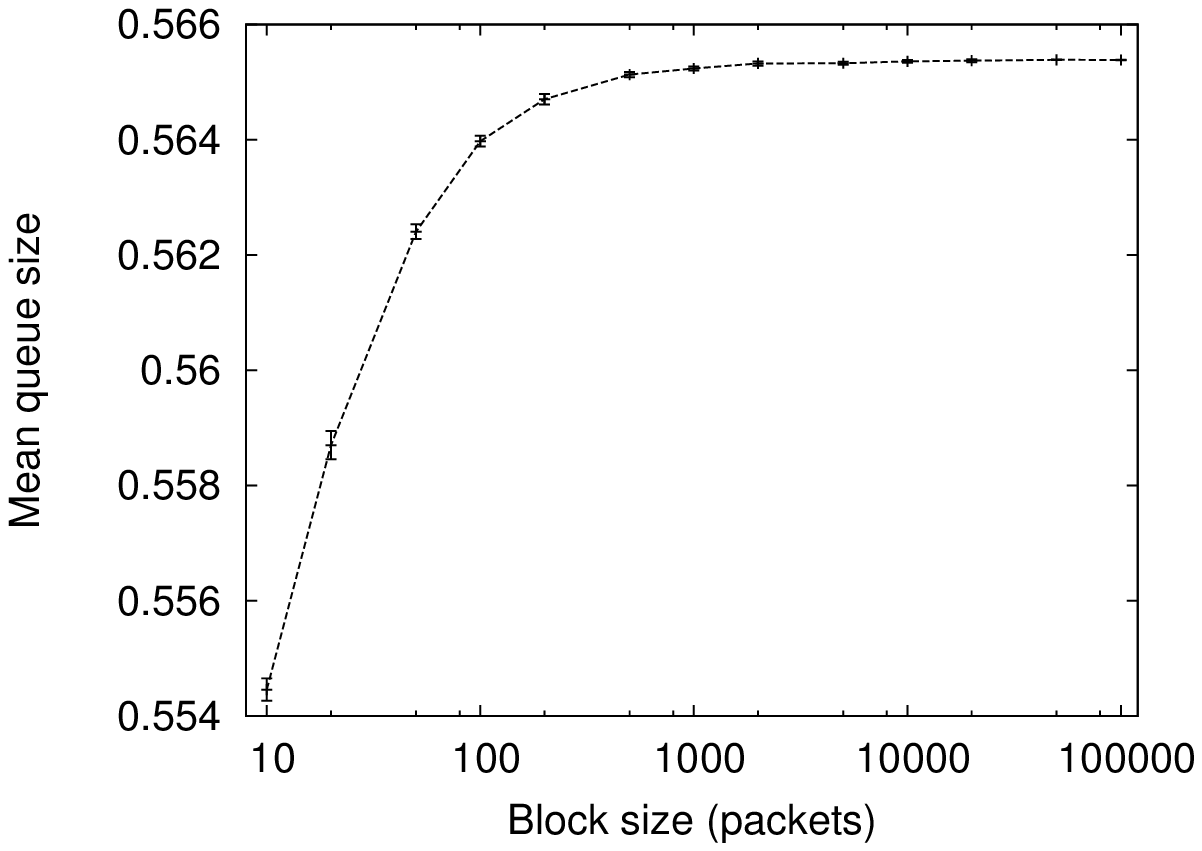} \\
\includegraphics[width=7.5cm]{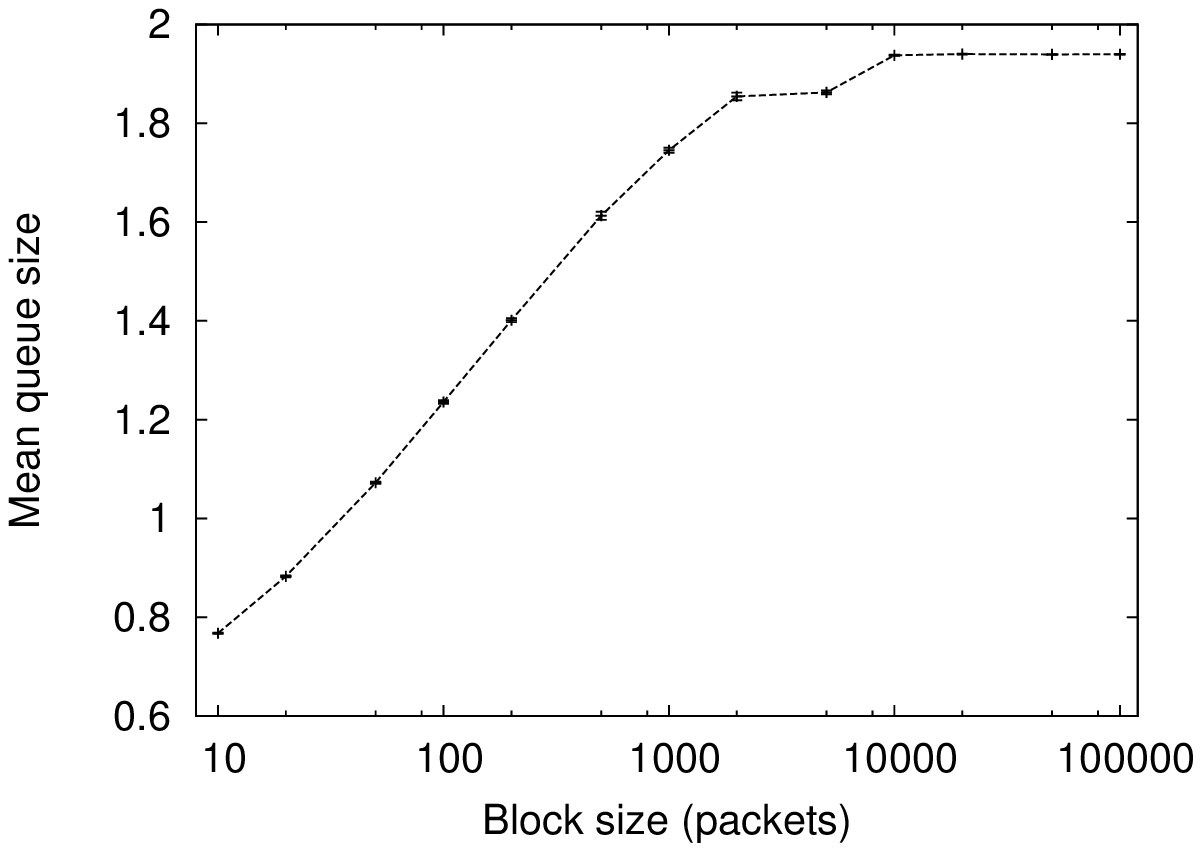}
\caption{Mean queue size versus blocksize for CAIDA data -- real data (top) and
simulated (bottom).}
\label{fig:oc48rearrange}
\end{center}
\end{figure}

Figure \ref{fig:oc48rearrange}
(top) shows this experiment on the CAIDA data.  Again there are ten repetitions
of each blocksize and the graph shows mean of the means and the standard deviation of
the means above and below.
Note the extremely small scale on the y axis.  
Even for very small  block sizes the variation
in the queuing performance is not great.  Beyond a block size of 1,000 the
correlation seems unimportant to the queuing performance and the resultant
mean queue size is the same to several decimal places.

By contrast, 
in figure \ref{fig:oc48rearrange} (bottom)
for the
LRD simulation almost all scales of correlation theoretically affect queuing
performance (within the bounds implied by the fact that only a finite sample of data
is used).  Here, the relationship between correlation and queuing is clearly shown.
Correlations of block sizes up to
10,000 packets are important to the queuing performance of the simulated data.
It is somewhat surprising that block sizes above this don't affect queuing
performance suggesting that in this simulated data set correlations above 10,000
packets were not important.  This is perhaps due to the relatively low Hurst 
parameter.

\begin{figure}[ht!]
\begin{center}
\includegraphics[width=7.5cm]{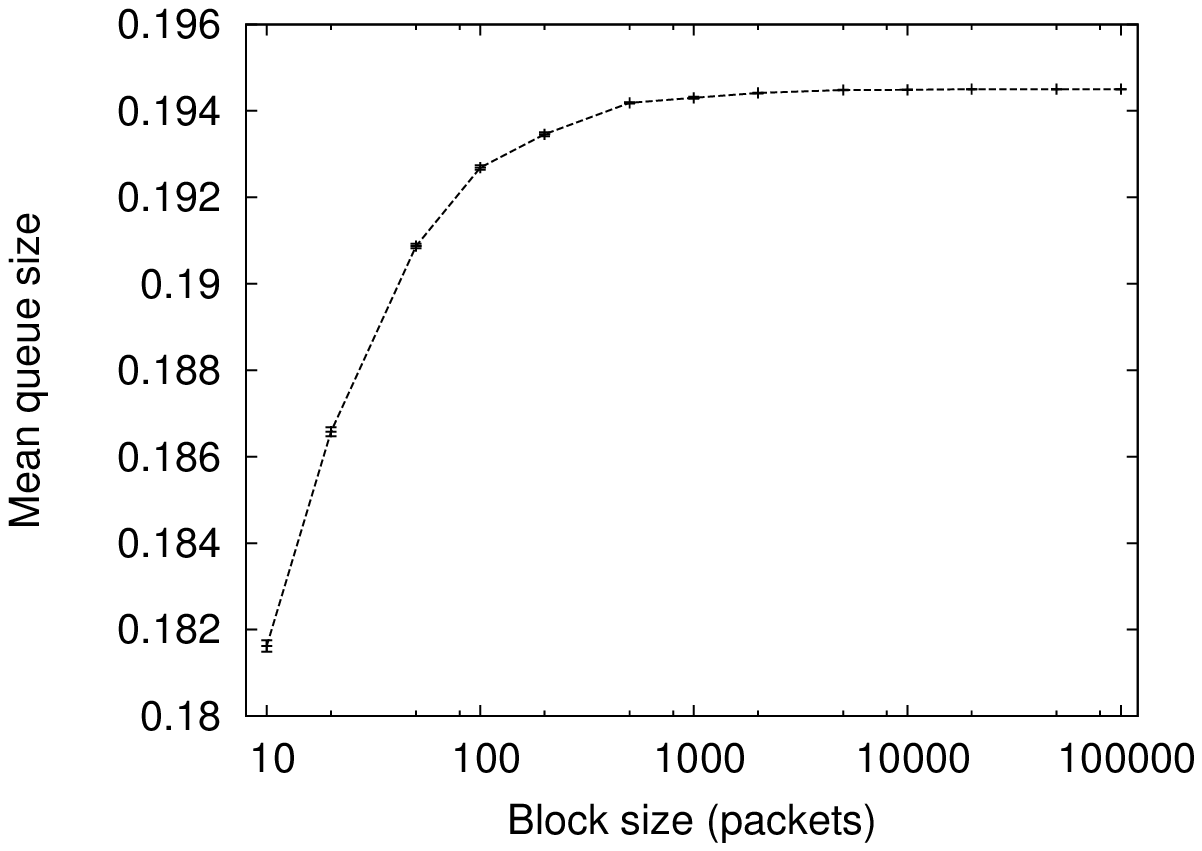} \\
\includegraphics[width=7.5cm]{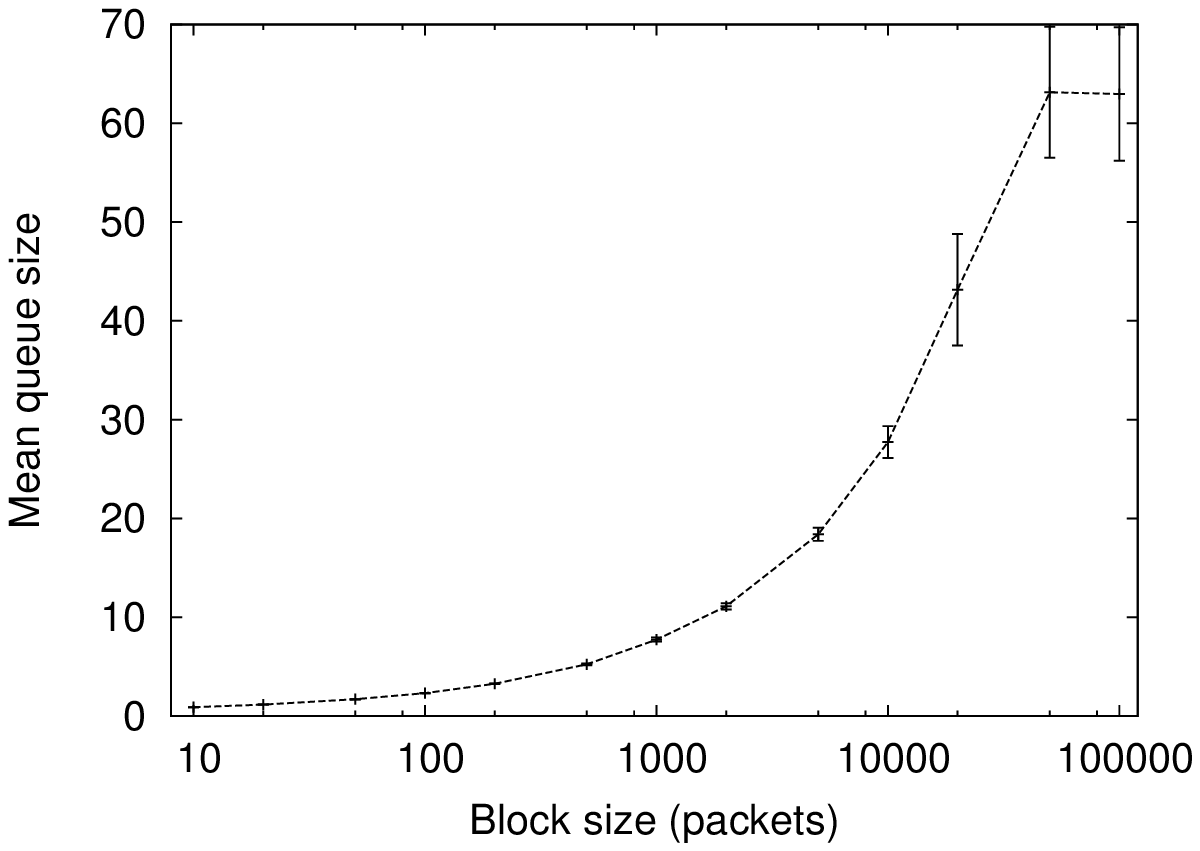}
\caption{Mean queue size versus blocksize for Bellcore data -- real data (top) and
simulated (bottom).}
\label{fig:bcrearrange}
\end{center}
\end{figure}

Figure \ref{fig:bcrearrange} shows the same experiment on the Bellcore data.  
This data has a higher Hurst parameter and the effect is more pronounced.  Again
the LRD method used drastically over predicts the level of queuing compared
to the real data when the same mean arrival rate and Hurst parameter is used.  
Again the same result is seen only a small difference in the amount of queuing
when the correlations are broken up.  The correlations of long time scale 
are unimportant
in the real data but important in the artificial data.  

\begin{figure}[ht!]
\begin{center}
\includegraphics[width=7.5cm]{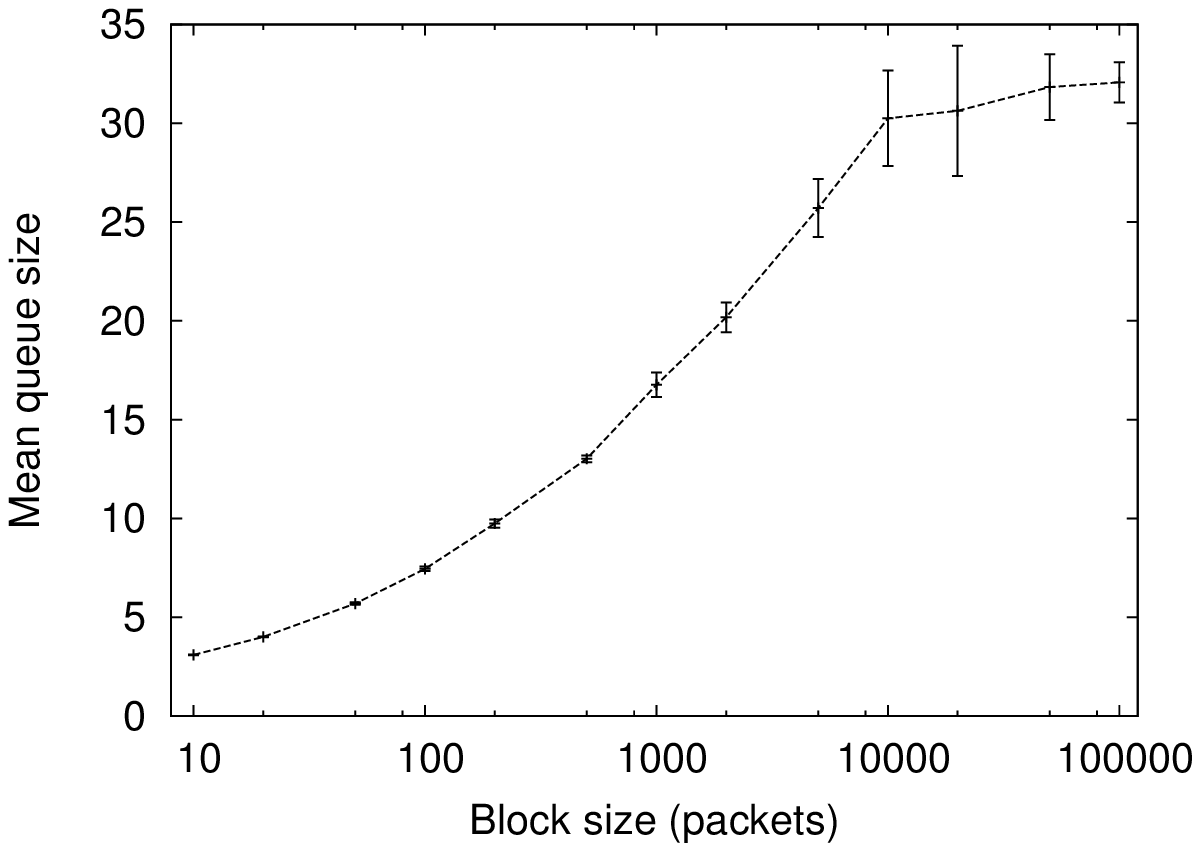} \\
\includegraphics[width=7.5cm]{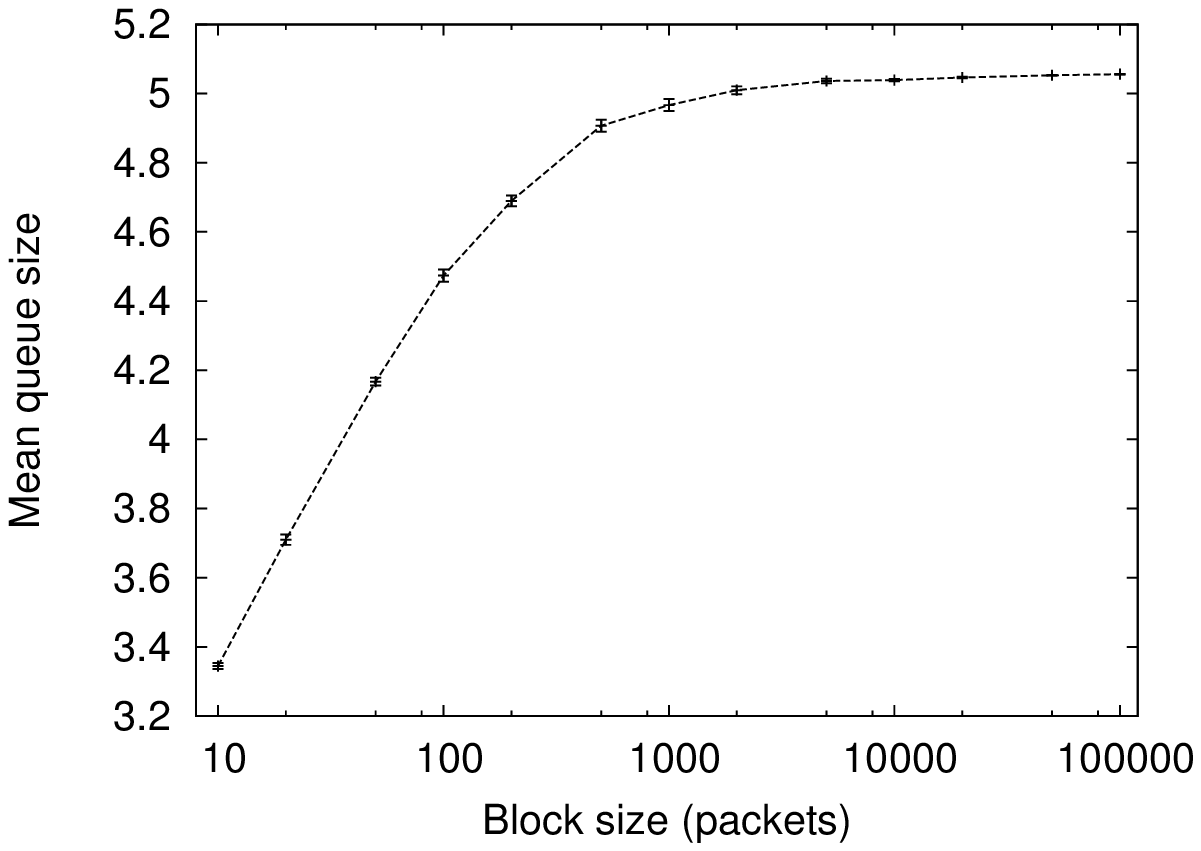}
\caption{Mean queue size versus blocksize for real data with high load -- 
Bellcore data (top) and CAIDA data (bottom)}
\label{fig:hitraffic}
\end{center}
\end{figure}

Again the question might be asked would the same conclusion hold true for
the real data if the bandwidth
used for the experiment were reduced. Figure \ref{fig:hitraffic} (top) shows
this for the Bellcore data with the bandwidth reduced so the queue occupancy is
extremely high (0.46).  Similarly
\ref{fig:hitraffic} (bottom) shows this for the CAIDA data (with an even higher
queue occupancy of 0.62).  These traffic levels would be extremely (unrealistically) 
high for a real network even at peak periods \cite{odlyzko2003}.

In the CAIDA data not much has changed except for the vertical axis scale.  In the
Bellcore data, however, with the much higher occupancy, more time scales are 
important.  This would, to some degree, be expected since at high occupancy queues
are more likely to persist for long periods of time.

The conclusion of this section is clear.  The claim that correlations over long scales
is important to queuing behaviour is not true of the CAIDA data and arguably true
of the Bellcore data only when the system occupancy is extremely high.

\section{Conclusions}

This paper criticises long-range dependence as a useful model for packet traffic.
Firstly, a theoretical problem with a class
of models used to simulate LRD is shown.  This class
of models predicts either no queuing or
an infinite expected queue length when fed into an infinite 
buffer.  At the very least,
experimenters should be aware of this problem to ensure that simulations
are not affected by it (the answer given by the simulation is a product of
the runtime of the simulation rather than a stable reflection of queuing 
performance).  
This effect is shown to be totally different to the queuing performance 
of real traffic.  The queuing behaviour of real network traffic queued
at a buffer which generates queues is totally different from that of
any single source based on a single heavy-tailed process which can
generate a queue at the buffer.

In the second part of the paper it is shown using simulated queuing on 
real traffic
that for real traffic traces long-range correlations are not important for
queuing behaviour except with extremely high traffic.  On the data sets
tests correlations over long time scales are not, in fact, important
for queuing behaviour with the arguable exception of the Bellcore trace
at extremely unrealistically high occupancy levels.  These 
results should be replicated on more traffic traces to work out how general
this conclusion is.

\bibliographystyle{abbrv}
\bibliography{nsrl_jccs2009}
\end{document}